\documentclass[twocolumns]{aa} 

\usepackage{rotating}
\usepackage{supertabular}
\usepackage{longtable}
\usepackage{graphicx}
\usepackage{natbib}
\usepackage{txfonts}
\usepackage{color}

\begin{document}
\newcommand{\fe}{[\ion{Fe}{ii}]}
\newcommand{\Ti}{[\ion{Ti}{ii}]}
\newcommand{\tii}{[\ion{Ti}{i}]}
\newcommand{\sii}{[\ion{S}{ii}]}
\newcommand{\oi}{[\ion{O}{i}]}
\newcommand{\oip}{\ion{O}{i}}
\newcommand{\pii}{[\ion{P}{ii}]}
\newcommand{\Ni}{[\ion{N}{i}]}
\newcommand{\Nii}{[\ion{N}{i}]}
\newcommand{\Nip}{\ion{N}{i}}
\newcommand{\caii}{\ion{Ca}{ii}}
\newcommand{\cai}{\ion{Ca}{i}}
\newcommand{\phoi}{[\ion{P}{i}]}
\newcommand{\cip}{\ion{C}{i}}
\newcommand{\he}{\ion{He}{i}}
\newcommand{\mgip}{\ion{Mg}{i}}
\newcommand{\mgiip}{\ion{Mg}{ii}}
\newcommand{\nai}{\ion{Na}{i}}
\newcommand{\brg}{Br\,$\gamma$}
\newcommand{\pab}{Pa\,$\beta$}
\newcommand{\fei}{\ion{Fe}{i}}
\newcommand{\feii}{\ion{Fe}{ii}}
\newcommand{\hei}{\ion{He}{i}}
\newcommand{\sip}{\ion{Si}{i}}
\newcommand{\mdot}{$\dot{M}_{jet}$}
\newcommand{\mjet}{$\dot{M}_{jet}$}
\newcommand{\macc}{$\dot{M}_{acc}$}
\newcommand{\mh}{$\dot{M}_{H_2}$}
\newcommand{\Ne}{n$_e$}
\newcommand{\h}{H$_2$}
\newcommand{\kms}{km\,s$^{-1}$}
\newcommand{\um}{$\mu$m}
\newcommand{\lam}{$\lambda$}
\newcommand{\msyr}{M$_{\odot}$\,yr$^{-1}$}
\hyphenation{mo-le-cu-lar pre-vious e-vi-den-ce di-ffe-rent pa-ra-me-ters ex-ten-ding a-vai-la-ble}

%

\title{The outburst of an embedded low-mass YSO in L1641
\thanks{Based on observations collected at the ESO/NTT (082.C-0264), at the REM telescope 
La Silla, Chile, and at the the Italian Telescopio Nazionale Galileo (TNG), operated on the island of La Palma by the Fundacion
Galileo Galilei of the INAF (Istituto Nazionale di Astrofisica)
.}}

\author{A. Caratti o Garatti \inst{1},
        R. Garcia Lopez \inst{1},
        A. Scholz \inst{1},
        T. Giannini \inst{2},
        J. Eisl\"{o}ffel \inst{3},
        B. Nisini \inst{2},
        F. Massi \inst{4},
        S. Antoniucci \inst{2},
        \and T.P. Ray \inst{1}
}
\institute{
Dublin Institute for Advanced Studies, 31 Fitzwilliam
Place, Dublin 2, Ireland \\ 
\email{alessio@cp.dias.ie}\\
\and
INAF - Osservatorio Astronomico di Roma, via Frascati 33, I-00040 Monte Porzio, Italy\\
\and
Th\"uringer Landessternwarte Tautenburg,
Sternwarte 5, D-07778 Tautenburg, Germany\\
\and
INAF - Osservatorio Astrofisico di Arcetri, Largo E. Fermi 5, I-50125 Firenze, Italy\\
}

   \date{Received ; accepted }

 
  \abstract
   {Strong outbursts in very young and embedded protostars are rare and not yet fully understood.
   They are believed to originate from an increase of the mass accretion rate ($\dot{M}_{acc}$) onto the source.}
   {We report the discovery of a strong outburst in a low-mass embedded young stellar object (YSO), namely 
   \object{2MASS-J05424848-0816347} or \object{[CTF93]216-2}, as well as its photometric and spectroscopic follow-up.}
   {Using near- to mid-IR photometry and NIR low-resolution spectroscopy, we monitor the outburst, deriving its magnitude,
   duration, as well as the enhanced accretion luminosity and mass accretion rate.}
   {\object{[CTF93]216-2} increased in brightness by $\sim$4.6, 4.0, 3.8, and 1.9\,mag in the $J$, $H$, $K_s$ bands and at 24\,$\mu$m,
   respectively, corresponding to an $L_{bol}$ increase of $\sim$20\,L$_{\sun}$. Its early spectrum, probably taken soon after the outburst,
   displays a steep almost featureless continuum, with strong CO band heads and H$_2$O broad-band absorption features, and
   Br$\gamma$ line in emission. A later spectrum reveals more absorption features, allowing us to estimate $T_\mathrm{eff}$$\sim$3200\,K,
   $M_*$$\sim$0.25\,M$_{\sun}$, and $\dot{M}_{acc}$$\sim$1.2$\times$10$^{-6}$\,M$_{\sun}$\,yr$^{-1}$.
   This makes it one of the lowest mass YSOs with a strong outburst so far discovered.}
  {}

\authorrunning{A.Caratti o Garatti et al.}
\titlerunning{Outburst of an embedded low-mass YSO}

   \maketitle
%

\section{Introduction}
\label{intro:sec}
Most of the stellar mass in low-mass YSOs is assembled within the first 10$^5$\,yr of their evolution~\citep[i.\,e. class\,0, 10$^4$\,yr, and Class\,I, 10$^5$\,yr: see e.\,g.,][]{lada84,andre93}.
During this stage, the YSO luminosity is thus expected to be dominated by accretion.
However, several studies, including the latest \textit{Spitzer Space Telescope} surveys~\citep[e.\,g.,][]{enoch,evans09}, 
have found that more than 50\% of the embedded YSOs have $L_{bol}$ and $\dot{M}_{acc}$ values considerably lower than those 
theoretically predicted~\citep[i.\,e. $\sim$2$\times$10$^{-6}$\,M$_\odot$\,yr$^{-1}$ for solar-mass YSOs;][]{shu,terebey}
and roughly of the same order of magnitude as the Classical T Tauri stars~\citep[CTTs; i.\,e. 10$^{-8} \le \dot{M}_{acc} \le$10$^{-6}$\,M$_\odot$\,yr$^{-1}$; e.\,g.,][]{white}.
Among several hypotheses, a likely explanation is that the mass accretion is \textit{episodic}, and the protostars with the lowest luminosities are those observed in quiescent accretion states~\citep{enoch,evans09,vorobyov}. 
Non-steady mass accretion is often observed in CTTs, such as EXors and FUors (lasting from a few months to several decades), 
in which $\dot{M}_{acc}$ increases by several orders of magnitude up to $\dot{M}_{acc}$$\sim$10$^{-4}$\,M$_\odot$\,yr$^{-1}$~\citep{hart}. 
It is thus reasonable to believe that a similar mechanism also exists in earlier and more embedded YSOs. 
Unfortunately, there is little direct observational evidence of outbursts in Class\,I YSOs, and so far, 
only a few clear cases have been detected~\citep[e.\,g. \object{V\,1647\,Ori} outbursts in 2003 and 2008, or \object{OO Ser} in 1995; see e.\,g.][]{fedele,kospal}. To reconcile theory with observations and improve the quality of the statistics, 
it is mandatory to detect and study these rare events.

With this aim, we started a long-term project to monitor the NIR flux and spectroscopic variability of embedded YSOs (mostly Class\,I and Flat sources) 
in nearby, young, and active star-forming regions (namely L\,1641, CrA, and the Serpens Molecular Cloud). 
This letter reports the outburst of an embedded YSO in L\,1641, namely \object{2MASS-J05424848-0816347}, hereafter \object{[CTF93]216-2}
($\alpha_{2000}$=05$^h$42$^m$48.48$^s$, $\delta_{2000}$=-08$\degr$16$\arcmin$34$\farcs$7).
This object was identified by our group as a low-mass embedded YSO (spectral type M, circumstellar $A_\mathrm{V}$$\sim$18\,mag) 
with a flat spectral index ($\alpha$=0.25, derived by fitting all the photometric data points from 2.2 to 24\,$\mu$m) 
and a bolometric luminosity of $\sim$1.9\,L$_{\sun}$ (Caratti o Garatti et al. in prep., hereafter CoG). It has been named
\object{[CTF93]216-2}, because it is relatively close to \object{[CTF93]216}~\citep[][]{chen93,chen94}, located about 38$\farcs$5 SW. Our Spitzer/IRAC images indicate that both YSOs have precessing jets, thus they could be part of a wide binary system ($\sim$17\,300\,AU,
assuming a distance $d$=450\,pc). During our recent survey in October 2010 with the robotic telescope REM (see Section\,\ref{observations:sec}), 
we detected for [CTF93]216-2 a brightness increase of several magnitudes with respect to the 2MASS images. We then compared our new images with the acquisition image in the $K_s$ band and the NIR spectrum of this source acquired in February 2009, discovering that the outburst was already in progress.
 

\section{Observations}
\label{observations:sec}

NIR spectra were obtained at the ESO-NTT with SofI~\citep{moor1} (on the 13 Feb.\,2009) and at the 3.5-m Italian telescope TNG
(13 Oct.\,2010) with NICS~\citep{baffa}, adopting the usual ABBA configuration.
The SofI spectrum was taken with the red grism (1.51-2.5\,$\mu$m), a 0\farcs6 slit $\mathcal R\sim$1000), and a
total integration time of 1800\,s. The full width half maximum (\textit{FWHM}) in the dispersion direction, measured from Gaussian fits to the OH sky lines, was $\sim$19\,$\AA$ ($\sim$260\,km\,s$^{-1}$) in the $K$ band.
The NICS low-resolution spectrum was acquired using a slit width of 1$\arcsec$ ($\mathcal R\sim$600) for the JH and HK grisms (1.15-1.75\,$\mu$m and 1.4-2.5\,$\mu$m, respectively) with a total integration time ($Int$) of 480 and 160\,s for the JH and HK grisms, respectively.
The measured \textit{FWHM} in the dispersion direction was $\sim$33\,$\AA$ ($\sim$600\,km\,s$^{-1}$) in the $K$ band.
Telluric and spectrophotometric standards were observed to correct for the atmospheric transmission and flux-calibrate the spectra.
The wavelength calibrations were performed using a xenon lamp in the infrared. 

Most of our $J$, $H$, and $K_s$ images were collected between the 10 October and the 6 November 2010 at the 60-cm robotic telescope 
REM~\citep[][ESO La Silla observatory]{zerbi} with the NIR camera REMIR~\citep{conconi} and a 150\,s total integration time per filter.
An additional $H$ band image was obtained with NICS (13 Oct.\,2010) with a 18\,s total integration time.
Additional $K_s$ photometry (5 Nov.\,2003) with the near-IR instrument  UIST~\citep{ramsay} was retrieved from the UKIRT data archive ($Int$=15\,s). 
Moreover, another photometric data point was derived from the $K_s$ band of the SofI $K_s$ acquisition image ($Int$=6\,s).
Early epoch photometry was retrieved from the Two-Micron All Sky Survey~\citep[2MASS;][$J$, $H$, $K_s$ band; November 1998]{2MASS}
and from the DEep Near-Infrared southern sky Survey~\citep[DENIS;][$J$ and $K_s$ bands; January 1997]{DENIS}.
All the raw data were reduced using 
\emph{IRAF}
packages, applying standard procedures for sky subtraction, dome
flat-fielding, and bad pixel and cosmic ray removal. 
Photometric calibration was obtained by means of photometric standard stars, except for the SofI $K_s$ acquisition image
and the UIST image, where 2MASS photometry of field stars was used.

Finally, we also used additional Spitzer/IRAC photometric data (3.6, 4.5, 5.8, and 8\,$\mu$m, obtained on the 8 Oct.\,2004), 
MIPS-24\,$\mu$m (taken on the 2 Apr.\,2005 and 27 Nov.\,2008), MIPS 70 and 160\,$\mu$m (2 Apr.\,2005), as well
as Spitzer/IRS low-resolution spectroscopy (5.2-39\,$\mu$m, obtained on the 12 Mar.\,2007), presented in CoG. 
Additional SCUBA/JCMT photometry (at 450 and 850\,$\mu$m) was taken from \cite{difrancesco}.

\section{Results}
\label{results:sec}
\begin{figure}
 \centering
\includegraphics [width=9.5cm] {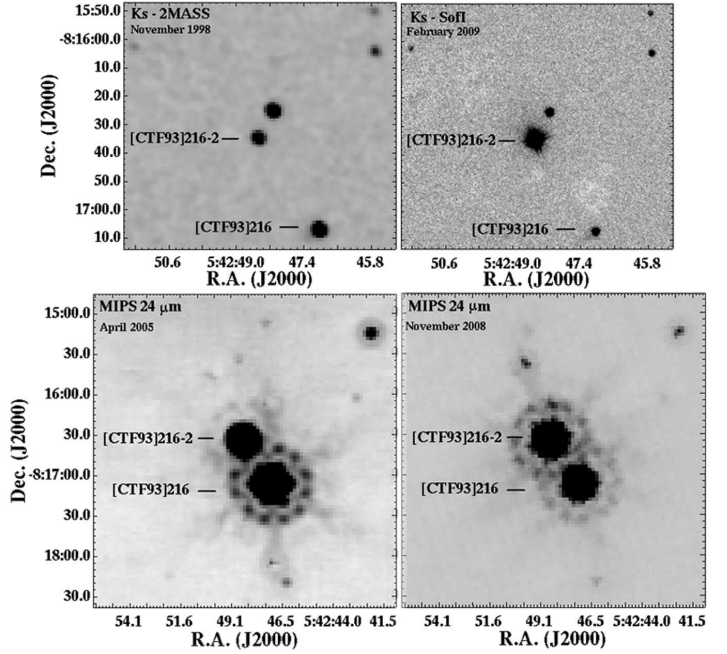}\\
  \caption{ $K_s$ ({\bf top panels}) and MIPS-24\,$\mu$m ({\bf bottom panels}) images of [CTF93]216-2 and its surroundings
before ({\bf left}) and after ({\bf right}) the outburst. The position of the YSO [CTF93]216 is also indicated.
\label{fig1:fig}}
\end{figure}

\begin{table}
\begin{scriptsize}
\caption{Photometry of [CTF93]216-2.}             
\label{phot:tab}      
\centering          
\begin{tabular}{cccccc}
\hline\hline\\ [-5pt]      
Date & JD         & $J$ & $H$  & Ks  & 24\,$\mu$m\\ 
(d.m.y)     & (2450000+) &  (mag) & (mag) & (mag) & (mag) \\
\hline                    
\hline\\[-5pt]
10.1.1997   &  0458.6 & 16.1$\pm$0.2   & $\cdots$       & 11.55$\pm$0.1   &    $\cdots$     \\
18.11.1998  &  1135.7 & 16.4$\pm$0.1   & 13.53$\pm$0.02 & 11.76$\pm$0.03  &    $\cdots$     \\ 
5.11.2003   &  2948.8 & $\cdots$       & $\cdots$       & 10.9$\pm$0.2    &    $\cdots$ \\  
2.4.2005    &  3462.8 & $\cdots$       & $\cdots$       & $\cdots$	 &    2.48$\pm$0.1   \\ 
12.3.2007   &  4172.0 & $\cdots$       & $\cdots$       & $\cdots$	 &    2.49$\pm$0.06\tablefootmark{a} \\
27.11.2008  &  4798.2 & $\cdots$       & $\cdots$       & $\cdots$	 &    0.58$\pm$0.04 \\
13.2.2009   &  4875.5 & $\cdots$       & 9.55$\pm$0.06\tablefootmark{b}  & 7.98$\pm$0.05  &    $\cdots$ \\
10.10.2010  &  5479.9 & 11.77$\pm$0.08 & 9.66$\pm$0.05  & 8.13$\pm$0.07  &    $\cdots$ \\
12.10.2010  &  5481.9 & 11.83$\pm$0.09 & 9.64$\pm$0.04  & 8.12$\pm$0.05  &    $\cdots$ \\
13.10.2010  &  5482.7 & $\cdots$       & 9.70$\pm$0.05  & $\cdots$       &    $\cdots$ \\
19.10.2010  &  5488.9 & 11.82$\pm$0.07 & 9.63$\pm$0.06  & 8.15$\pm$0.07  &    $\cdots$ \\
28.10.2010  &  5497.8 & 12.01$\pm$0.1  & 9.76$\pm$0.09  & 8.25$\pm$0.1   &    $\cdots$ \\
06.11.2010  &  5506.6 & 12.00$\pm$0.07 & 9.70$\pm$0.06  & 8.24$\pm$0.08  &    $\cdots$ \\
\hline                  
\end{tabular}
\tablefoot{
\tablefoottext{a} {Photometry derived from the Spitzer-IRS flux-calibrated spectrum.}
\tablefoottext{b} {Photometry derived from the SofI flux-calibrated spectrum.}
}
\end{scriptsize}
\end{table}
\begin{figure}
 \centering
\includegraphics [width=9cm] {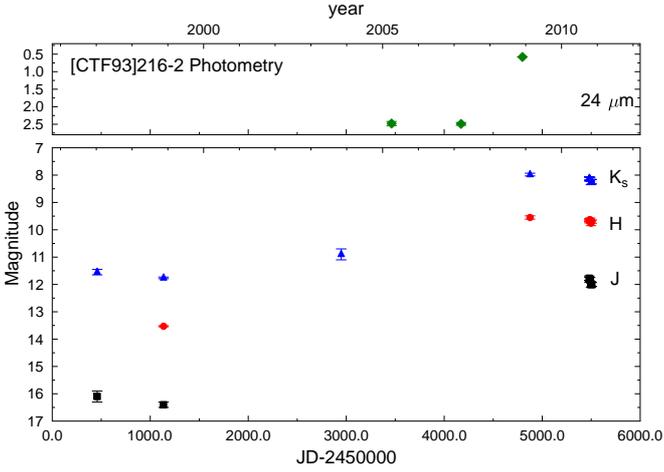}\\
  \caption{ [CTF93]216-2 lightcurves. $J$, $H$, $K_s$, and 24\,$\mu$m data are represented as black squares,
red dots, blue triangles, and green diamonds, respectively.
\label{fig2:fig}}
\end{figure}

\subsection{Photometry}

\begin{figure}
 \centering
\includegraphics [width=8cm] {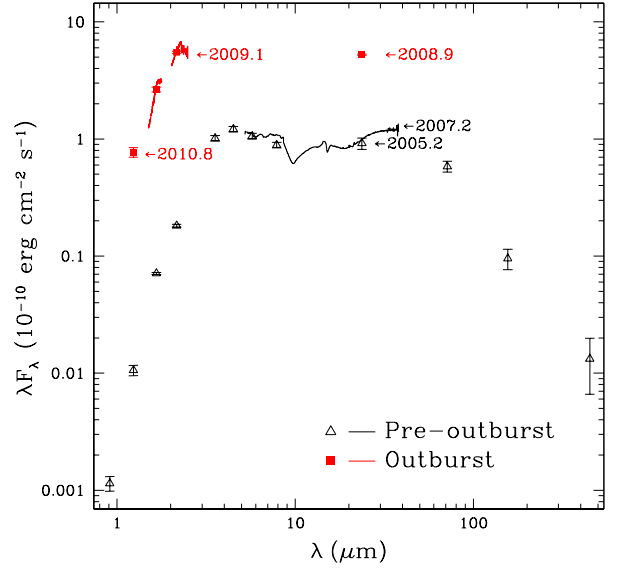}\\
  \caption{ Spectral energy distribution of [CTF93]216-2. \textit{Pre-outburst data}: black triangles (photometry),
and black line (Spitzer-IRS mid-IR spectrum). \textit{Outburst data}: red squares (photometry), 
and red line (SofI NIR spectrum).
\label{fig3:fig}}
\end{figure}

In Figure~\ref{fig1:fig}, pre- and outburst images of [CTF93]216-2 
in the $K_s$ band (top panels) and the Spitzer/MIPS 24\,$\mu$m band (bottom panels) are shown,
clearly indicating an increase in the object brightness. 
Lightcurves in the $J$, $H$, $K_s$ bands, and at 24\,$\mu$m are shown in Figure~\ref{fig2:fig},
and the corresponding photometric data points along with their uncertainties are reported in Table~\ref{phot:tab}.
We stress that the photometric data points at 24\,$\mu$m (on the 12 Mar.\,2007) and in the $H$
band (on the 13 Feb.\,2009) were derived from the Spitzer/IRS and from the NTT/SofI flux-calibrated spectra, respectively.
The 24\,$\mu$m data points (Fig.~\ref{fig2:fig}, upper panel) give us some constraints on the date of the outburst,
i.\,e. after March 2007 and before November 2008. 
Pre-outburst photometry from DENIS, 2MASS, and the UIST $K_s$ images indicates that the object is variable
(in both $J$ and $K_s$ bands, we have just one 2MASS data point for the $H$ band), as is often the case for YSOs~\citep{carpenter}.
Comparing 2MASS (November 1998) with our earliest outburst SofI photometry (February 2009), we measure the amplitudes
$\Delta K_s$$\sim$3.8\,mag, and $\Delta H$$\sim$4\,mag, corresponding to a flux increase $\Delta F$ by a factor of $\sim$33 and 40, 
respectively.
Considering our earliest post-burst photometric point in the $J$ band (October 2010), we derive $\Delta J$$\sim$4.6\,mag (or 
$\Delta$F$\sim$69), whereas the $\Delta mag$ at 24\,$\mu$m is $\sim$1.9\,mag ($\Delta$F$\sim$6).
Our data do not allow us to determine the exact date of the peak, but it is clear that the NIR lightcurves
are still close to the maximum, and that this plateau phase has lasted at least 2 years,
with a small decrement in the $H$ and $K_s$ bands of about 0.1\,mag between February 2009 and October 2010.
Moreover, our latest photometry (November 2010) seems to indicate that the rate of decrease, between October and November 2010, 
has increased to $\sim$0.2\,mag/month in the $J$ band and $\sim$0.1\,mag/month in the $H$ and $K_s$ bands.
Pre- and outburst spectral energy distributions (SEDs) are shown in Figure~\ref{fig3:fig}, where all available photometric and spectroscopic
data (from this work and from CoG) are reported. 
We note that both the spectral index and bolometric luminosity have changed, as already reported in other
YSO outbursts~\citep[see e.\,g. \object{OO Ser},][]{kospal}.
During the outburst, the YSO SED has become bluer and flatter, and the $\alpha$ value, computed between 2.2 and 24\,$\mu$m, varied from 0.69 to -0.04.
To compute $L_{bol}$, and estimate the outburst $\Delta mag$ in the Spitzer/IRAC photometry, we fit the dereddened 
$\Delta mag$ amplitude from 1.25 to 24\,$\mu$m by means of a power law, obtaining $\Delta m$(3.6\,$\mu$m)=2.9\,mag, $\Delta m$(4.5\,$\mu$m)=2.6\,mag, 
$\Delta m$(5.8\,$\mu$m)=2.3\,mag, and $\Delta m$(8\,$\mu$m)=2.1\,mag. 
We also assume that the SED flux experienced no significant increase for $\lambda>$24$\mu$m. 
As a result, we estimate an $L_{bol}$  value of $\sim$22\,L$_{\sun}$ during the outburst,
i.\,e., $\Delta L_{bol} \sim$20\,L$_{\sun}$ with respect to the pre-outburst state.

\subsection{Spectroscopy}

\begin{figure*}
 \centering
\includegraphics [width=18cm] {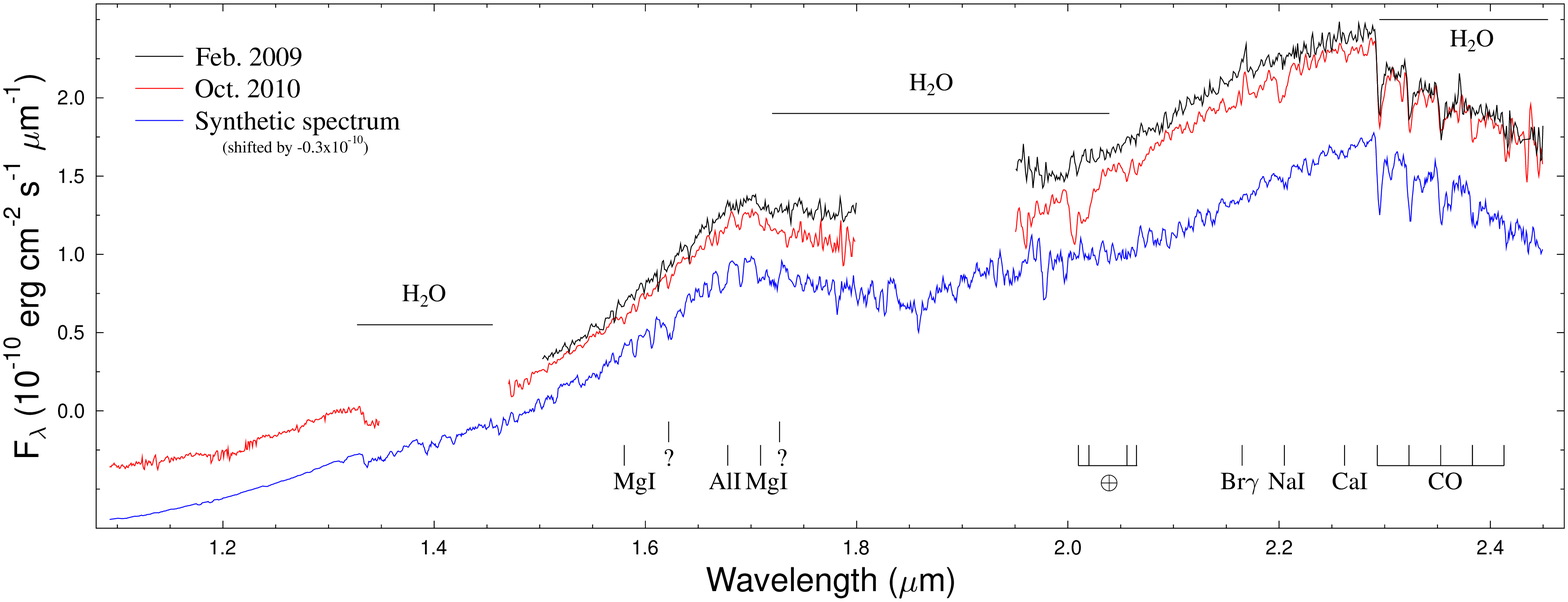}\\
\caption{Flux calibrated NIR spectra of [CTF93]216-2: SofI spectrum (black), 
and NICS spectrum (red). For comparison, we show an AMES-DUSTY
model spectrum for effective temperature of 3200\,K and log g=4.0~\citep{allard}.
The DUSTY spectrum has been shifted by -0.3$\times$10$^{-10}$\,erg\,s$^{-1}$\,cm$^{-2}$\,$\mu$m$^{-1}$ for clarity.
The detected features are also labelled.
\label{fig4:fig}}
\end{figure*}

Our SofI and NICS low resolution spectra (taken in February 2009 and October 2010, respectively)
are shown in Figure~\ref{fig4:fig} (in black and red). 

The SofI spectrum shows an almost featureless continuum, with strong veiling and IR excess.
The absorption water band between 1.7 and 2.1\,$\mu$m, typical of
M spectral types with a $T_\mathrm{eff}\le$3800\,K, is clearly visible. 
In addition, deep CO band-head absorption lines are detected at 2.29\,$\mu$m (2-0), 2.32\,$\mu$m (3-1), and 2.35\,$\mu$m (4-2). 
Finally, Br$\gamma$ emission, usually associated with accretion, is observed in our SofI spectrum. 
The measured Br$\gamma$ flux is 9.6($\pm$1.8)$\times$10$^{-14}$\,erg\,s$^{-1}$\,cm$^{-2}$ (EW$\sim$-3.6\,$\AA$), 
peaking at 2.166\,$\mu$m.

The flux of the NICS spectrum, taken 607\,days later, shows a slight decrease. The continuum still
has a similar shape, but the absorption H$_2$O band depth is more prominent.
The EW of the CO lines slightly increases by $\sim$1\,$\AA$ (9.8 to 10.6\,$\AA$ and 7.2 to 8.2\,$\AA$, for
the 2-0 and 3-1 band heads, respectively). More CO lines at 2.38\,$\mu$m (5-3), and 2.41\,$\mu$m (6-4) are detected.
In addition, a few more narrow-band absorption features become visible on the continuum, i.\,e. \ion{Ca}{i} (2.26\,$\mu$m),
the \ion{Na}{i} doublet (2.20-2.21\,$\mu$m), \ion{Mg}{i} (1.59-1.71\,$\mu$m), and the \ion{Al}{i} doublet (1.67-1.68\,$\mu$m).
No relevant features are detected in the $J$ band, except for the absorption H$_2$O band between 1.3 and 1.5\,$\mu$m.
The Br$\gamma$ line is not resolved, peaking at 2.166\,$\mu$m. 
The measured flux is 9($\pm$2)$\times$10$^{-14}$\,erg\,s$^{-1}$\,cm$^{-2}$ (EW$\sim$-3.7\,$\AA$).
We note that no signature of any emission line (such as H$_2$ or [\ion{Fe}{ii}]) from the jet has been detected on-source
in both spectra.

To more tightly constrain the YSO spectral type, we separately compared limited
regions of our spectra (in the $H$ and $K$ bands) to model spectra. Veiling
and scattering (not taken into account by our fit) are expected to 
smoothly and slowly vary as a function of wavelength and thus, they
should not affect the spectrum in individual bands significantly~\citep[e.\,g.,][]{scholz10}. 
For our comparison, we used a series of AMES-DUSTY
model spectra~\citep{allard} with $T_\mathrm{eff}$ ranging from 2500
to 3900\,K and log\,g=4.0, as expected for young stars. We varied
$T_\mathrm{eff}$ to obtain a consistent solution for the two bands, using
the $A_V$=18\,mag derived by CoG. We find a reasonable match
between observed and model spectra for $T_\mathrm{eff}$=3200$\pm$200\,K.
As an illustration, the model for 3200\,K is shown in Fig.~\ref{fig4:fig}
(blue spectrum), shifted by -0.3$\times$10$^{-10}$\,erg\,s$^{-1}$\,cm$^{-2}$\,$\mu$m$^{-1}$ for clarity.


\section{Discussion }
\label{discussion:sec}

Our discovery of the [CTF93]216-2 outburst gives us a rare opportunity 
to study boosted accretion in young embedded protostars, probing whether 
episodic mass accretion can reconcile the low accretion rates
observed in young embedded protostars with the theoretical predictions.

The Br$\gamma$ luminosity, corrected for the circumstellar extinction, 
is often used to derive an estimate of the accretion luminosity~\citep[see e.\,g.][]{muzerolle98,natta06}.
Assuming that $A_\mathrm{V}$=18$\pm$3\,mag and $d$=450\,pc, we derive $L(Br\gamma)$$\sim$3.5(0.9)$\times$10$^{-3}$\,L$_{\sun}$ and, 
from the \citet{muzerolle98} relationship, we obtain $L_{acc}$$\sim$22$\pm$8\,L$_{\sun}$, which is
close to the derived $\Delta L_{bol}$$\sim$20\,L$_{\sun}$ and indicates that $L_{acc} \sim \Delta L_{bol}$.

On the other hand, $\dot{M}_{acc}$ can be derived from 
$\dot{M}_{acc} = L_{acc} \times 1.25 R_* / G M_*$~\citep[][]{gullbring},
where $M_*$ and $R_*$ are the stellar mass and radius, and the number 1.25 is derived by assuming
a value of 5$R_*$ for the inner radius of the accretion disk.

Assuming that $T_\mathrm{eff}$=3200$\pm$200\,K and $L_{*}$$\sim$1.9$\pm$0.1\,L$_{\sun}$, from
\citet{siess} models, we obtain $M_*$=0.24$\pm$0.04\,M$_{\sun}$ and $R_*$=4.4$\pm$0.4\,R$_{\sun}$,
which gives a mass accretion rate of $\sim$1.1-1.3$\times$10$^{-6}$\,M$_{\sun}$\,yr$^{-1}$.
While $M_*$ is well constrained by $T_\mathrm{eff}$, the radius $R_*$ depends on $L_{*}$. Thus this last should be considered 
as an upper limit, since pre-outburst $L_{acc}$ is unknown. 
However, even a 50\% decrease in $L_{*}$ would affect $R_*$ and thus $\dot{M}_{acc}$ of just 20\%.

Compared to typical accretion rates of Class\,I YSOs with similar 
masses~(i.\,e. 10$^{-8}$\,M$_{\sun}$\,yr$^{-1}$; e.\,g., CoG; \citealt{scholz10,white}),
the derived $\dot{M}_{acc}$ is about two orders of magnitude higher. 
This value is about an order of magnitude lower than
what some episodic-accretion models predict for these early YSO bursts~\citep[i\,.e 10$^{-5}$\,M$_{\sun}$\,yr$^{-1}$; e.\,g.][]{vorobyov},
which should resemble the more evolved FUor counterparts.
On the other hand, we note that the magnitude of luminosity change during this outburst 
($\sim$10) is similar to those of \object{V\,1647\,Ori} and \object{OO Ser} ($\sim$8). 
Our object would probably differ from the others in its pre-outburst $L_{bol}$, which is about two times lower, because of its lower mass. 
Moreover, as already noted in previous cases~\citep[e.\,g.][]{hodapp,fedele,kospal}, the
[CTF93]216-2 outburst shows similarities with both FUor and EXor events, i.\,e. 
a NIR featureless FUor spectrum with strong absorption CO band heads, but with Br$\gamma$ line in emission as in EXors.
The event duration (if confirmed by additional observations) and amplitude are in-between those of EXors and FUors, thus,
as noted by other authors~\citep{gibb,fedele} EXors and FUors might not be distinct categories of eruptive events, but instead part
of a continuum of outburst events.





\begin{acknowledgements}
ACG was supported by the Science Foundation of Ireland, grant 07/RFP/PHYF790. 
We would like to thank Dino Fugazza and Francesca Ghinassi for supporting the REM and TNG observations.

\end{acknowledgements}

\bibliographystyle{aa}
\bibliography{ref}


\end{document}